
\magnification=1200
\baselineskip=12pt
\def\DM{\displaystyle}
\def\Eb{ {\phantom{p}} }
\def\DEF{\equiv}
\def\EQNO{\global\advance\count255 by 1 \eqno (\number\count255) }
\def\sv#1{\global\long\xdef#1{\number\count255}}
\def\ii{\imath}
\def\jj{\jmath}
\def\G#1#2{\gamma^{[{#1}_{{#2}}]}}

\hoffset=0.5truein
\hsize=6truein
\count255=0

\settabs 20 \columns
\+&&&&&&&&&&&&&&& McGill/94--14 \cr
\+&&&&&&&&&&&&&&& UM-TH-94-29 \cr
\+&&&&&&&&&&&&&&& hep-ph/9408341\cr
\+&&&&&&&&&&&&&&& August 17, 1994 \cr

\vskip2truecm

\centerline{\bf Renormalization of four-quark operators, effective theory}
\centerline{\bf and the role of evanescent operators}

\vskip1.5truecm

\centerline { Kassa Adel }
\medskip

\centerline { \sl Department of Physics, McGill University }
\smallskip

\centerline { \sl Montreal, QUE H3A 2T8, Canada}
\bigskip

\centerline {York-Peng Yao}
\medskip

\centerline { \sl Randall Laboratory of Physics, University of
Michigan }
\smallskip

\centerline { \sl Ann Arbor, MI 48109 U. S. A. }

\vskip2truecm

\centerline{\bf Abstract}
\bigskip

We present, in the context of dimensional regularization,
a prescription to renormalize Feynman diagrams
with an arbitrary number of external fermions. This prescription,
which is based on the original t'Hooft-Veltman proposal
to keep external particles in four dimensions,
is particularly useful to define the 'renormalization' (in the context of
effective Lagrangian) of
physical four-quark operators without introducing any
evanescent operator. The results obtained
for $b\rightarrow s$ processes agree with those from the so-called
naive prescription, but disagree
with the ones with the introduction
of evanescent operators in a renormalization group analysis.
We also present an explicit two loop calculation of the mixing of
the evanescent operators with the physical dimension five operators
for the same processes.  Particular attention is paid to
the unboundedness nature of such mixing and how a formal finite
transformation is effected to decouple.
The inevitable mass dependence of one of these schemes in the
literature is pointed out as the cause for the difference
mentioned.
\bigskip
\noindent
PACS numbers(s): 11.10.Jj, 12.38.Bx, 13.40.Hq
\vfill \eject
\baselineskip=20pt
\bigskip\noindent
{\bf 1. Introduction}
\medskip
There are at least three independent complete calculations of
$b\rightarrow s$ processes which disagree with each other Refs.[1,2,3].
Although the difference is numerically insignificant for $b \rightarrow
s \gamma $ (within $2\% $), it is theoreticaly important to obtain
the correct Wilson coefficients which contain all the heavy top
and W-boson effects.  The disagreement comes mainly from the
introduction of {\it evanescent } four quark operators in Ref.[2,3].  These
were not included in the analysis of Ref.[1], the contributions of which
were subsequently
given in Ref.[4], which brought Ref.[1] and Ref.[3] into agreement.
In this article, we shall attempt to give
some partial answer to where and when one should introduce these
evanescent operators.  We shall argue that if one is interested in
constructing an effective Lagrangian to a particular {\it finite } order in
the strong QCD coupling constant $g_s$, based on direct Feynman diagram
evaluation of processes, then inclusion or exclusion of these
operators is an arbitrary choice.  We shall present a scheme for
this purpose,
in which there is no need to include any evanescent operators.  It
is based on the original t'Hooft-Veltman (HV) prescription to keep the
external particles in four dimensions.  However, if one is to perform
a renormalization group analysis, then one has to be concerned also
with the closure of the operator basis under multiplicative renormalization
and the
light mass independence aspect of the Wilson coefficients.  Although
we have not been able to pinpoint, we hope that the following
presentation of our concern with respect to these last two items
may spur further investigation.

\bigskip
Before we proceed any further into the technical details, we now
give a brief exposition of the issues, so that an interested reader
who may not be a practitioner can have an overview.
\bigskip
A singular development in theoretical physics in the past twenty years
or so has to do with the successes of formalisms in short distance
analysis.  In many physical situations, it has been shown that scales
far different from the ones under consideration can be, very loosely speaking,
'integrated out' and their residual effects can be parameterized in
a series of terms made of effective coefficients and local products
of low excitations.  Particularly in particle physics, let $\psi $ be the
fields which are being prepared and detected experimentally.  One is
interested here in expanding a theory capable of more detailed
description at a higher energy scale $\Lambda $ into terms schematically as
$$L^{full}=L^{low \ energy}+{1\over \Lambda ^2}\sum _i \bar C_iO_i+\cdot
\cdot \cdot . \eqno (i)$$
If $L^{full}$ is a renormalizable theory, then $L^{low \ energy}$ will
include all the renormalizable terms after such an "integrating out'
of the heavy particles.  $\bar C_i$ are the Wilson coefficients, which are
functions of $\log \Lambda $ and perhaps some other variables and $O_i $
are made of local functions of $\psi $.  We have put a bar over $C_i$ to
indicate that their $1\over \Lambda ^2$ factor has been taken out.
This is a well-defined construct under the name of effective
Lagrangian.
\bigskip
Let us discuss some aspects of the actual construction of an effective
Lagrangian.  Specifically, we consider only the case when $L^{full}$ is
known and renormalizable.  One can perform a loop expansion with it,
from which the right hand side of Eq.(i) can be extracted.  The most
systematic way to do this to a particular order is by using
Zimmermann oversubtraction identities to rearrange the integrand
for any
diagram in which there is at least one heavy internal particle.  Terms
are organized or discarded according to dimensional counting.  Since
the role of the identities is merely to add and to subtract the same
terms to a given integrand to allow power counting down to parts of
it, there are many possibilities.  The common tread is an expansion
in external momenta and internal masses of these subdiagrams
relative to $\Lambda $.  A point which needs to be emphasized, however, is
that after the required wave function and parameter renormalizations
inherent in $L^{full}$, all these parts are finite quantities.  If
the regulator used for the said  renormalizations is dimensional
continuation, then the limit to go to four dimensions can be taken at
this stage and each term remains finite.  The advantage here is that
there are fewer operators $O_i$, which need to be introduced. They are called
dimensionally physical.  On the other hand, one may choose to continue
working slightly away from four dimensions after each finite order,
because the algebraic aspect of the operators may be more systematic.
This comes about because the mechanics of constructing an effective
Lagrangian is such that
one needs to reinsert the lower order pieces into
a diagram in order to obtain from it the higher order pieces.  In
so doing, one may find it convenient to introduce  extra operators,
which are termed evanescent because they vanish
formally in the four dimension limit.  These are some of the common
choices one is free to make.  They all should produce exactly the
same S-matrix elements, which are  dictated by $L^{full}$, to the
accuracy in any inverse power of $\Lambda $ demanded by us.

\bigskip
We now turn to the determination of $\bar C_i$ in infinite partial sums.
As it is well-known, when we include QCD, we should view $\bar C_i$ as
functions of $g_s^2$ and $g_s^2\log \Lambda ^2$, and perhaps of other
variables also as we shall see.  Because $g_s^2\log \Lambda ^2$ is not
small, an improved perturbation expansion is to sum infinite series
in leading powers, subleading powers, etc. of it.  The tool for this is
the renormalization group analysis.  Essential to the procedure is the
introduction of a dimensionful parameter $\mu $, which in dimensional
continuation is associated with the regularization program.  As our
earlier discussion reveals, terms such as $\sum_i \bar C_i O_i $ are finite
quantities, upon expressing them in renormalized fields, masses and
coupling constants.  It should also be noted that once a choice is
made of the Zimmermann rearrangement, there ensues a particular
renormalization prescription for $O_i .$  Schematically, using the
{\it closure} property
$$O_i=Z_{ij}O_j^{bare},\eqno (ii)$$
we write
$$\mu {d\over d\mu}O_i+\gamma _{ij}O_j=0, \eqno (iii)$$
with
$$\gamma _{ij}=-(\mu {d\over d\mu }Zik)Z^{-1}_{kj}, \eqno (iv)$$
and then it follows that
$$\mu {d\over d\mu}\bar C_i=\bar C_j\gamma _{ji}.  \eqno (v)$$
It is obvious that a preferential scheme is one which is {\it mass-
independent}, such that $\gamma _{ij}$ are functions only of $g_s (\mu )$
and $\bar C_i$ only of $g_s(\mu)$ and $\log (\Lambda /\mu )$.  This allows one
to perform infinite sums over leading logarithm, subleading logarithm,
etc more conveniently.
\bigskip
We shall develop this program in much greater details in the sections
to come.  Thus, the article is organized as follows.  In section 2, we review
again the general formalism for calculating directly the Wilson coefficients
from Feynman diagrams and explain how the effective theory is derived
from the full theory
(Standard Model).  We also discuss the scheme-dependence of various
renormalized quantities in the effective theory.  In section 3, we present
the renormalization group equations (RGE) for the Green's functions and
the Wilson coefficients and discuss the meaning of the leading logarithm
approximation (LLA) and the next-to-leading logarithm approximation
(NLLA).  In section 4, we present a new prescription to obtain
the physical four quark operators without introducing any evanescent
operators.  This scheme was implicitly used in Ref.[1] and therefore,
if adhered to for (RGE) analysis, all the results in the LLA will be
the same as
those of Ref.[1].  In section 5, we follow others in introducing
evanescent operators and perform a two loop calculation of the mixing
of these operators with the physical dimension five operators.  We shall
discuss the potential difficulty in dealing with unbounded infinite matrices.
We shall give the similarity transformation which formally decouples these
two sets of operators at any running scale $\mu $.  It is the same
transformation which makes the matrix
elements of these evanescent operators to go to $b \rightarrow
s \gamma, \ s G$ vanish at one loop.  In section 6, we comment on the results
and give conclusions.

\bigskip\noindent
{\bf 2. Effective theory}
\medskip

Consider the process $b\rightarrow s \gamma$ in the Standard Model ($SM$).
The diagrams contributing to this process to lowest order are shown in
Fig. 1.  Suppose we want to calculate the value of each
$\underline{\rm renormalized}$
diagram in the limit $m_t\sim m_w \gg p_{ext},m_b,m_s,m_u,m_c$ to an accuracy
${\cal {O}}
({1/ m_w^2})$, where $p_{ext}$ represents here the momemta of the
external particles.  We use the first diagram, redrawn in Fig. 2, to
explain how the effective theory is built to this order.
We do this by partitioning the diagram into several pieces to extract
individually the $1/m_w^2$ effects,
an approach which is explained in details in Ref. [5].
Let $\Gamma$ denote the unrenormalized
diagram on the left of Fig. 2 and let us define:
$$\eqalign{
& \tau_{\gamma}^{(\epsilon)}  \DEF {\rm  Pole\ Part\ of\ }\gamma, \cr
& \tau_{\gamma}^{(r)}  \DEF     \gamma (p=0)
            + p {\partial \over \partial p} \gamma(p=0)  + \dots
        + {1\over r!}\,p^r {\partial ^r \over \partial p^r} \gamma(p=0). \cr
}\EQNO $$
where $p$ represents the external momenta of the graph (or subgraph)
$\gamma$, and $\epsilon = n-4$, $n$ being the number of space-time
dimensions. The renormalized $\Gamma$ is obtained
by subtracting $\Gamma$ at $p_{ext} = 0$,
$$\Gamma^{ren} = \left(1-\tau_{\Gamma}^{(0)} \right) \Gamma .\EQNO\sv{\xx} $$
Let $\gamma_0$ be the tree subgraph of $\Gamma$ which contains the $W$-boson
line as illustrated in Fig. 3. We can now rewrite Eq. (\xx) as:
$$\eqalign{
\Gamma ^{(ren)} =\ \ & \left( 1 - \tau_{\Gamma}^{(2)} \right) \,
                          \,\tau_{\gamma_{0}}^{(0)} \, \Gamma  \cr
                + & \left( \tau_{\Gamma}^{(2)} - \tau_{\Gamma}^{(0)} \right)\,
			\Gamma \cr
                + & \left( 1 - \tau_{\Gamma}^{(2)} \right) \,
                 \left(\, 1 - \tau_{\gamma_{0}}^{(0)} \right) \, \Gamma . \cr}
\EQNO\sv{\xfirst}  $$
The first and second terms on the right in Eq. (\xfirst) correspond to the
first and second diagrams on the right in Fig. 2.
The last term in Eq. (\xfirst) is of order $1/m_w^4$ and is discarded.
Eq. (\xfirst) is nothing but an algebraic rearrangement, first used
by Zimmermann [6]. We would like to point out that every term is finite
and well defined; in particular when computing the quantities
in Eq. (\xfirst),
the Dirac algebra can be done with no
ambiguity in four dimensions .
This rearrangement or partitioning of graphs tells us how to compute
the heavy effects in the full theory. It also generates
an effective theory, order by order in perturbation theory.

\medskip

In the language of effective theory, the graphs on the right in Fig. 2
are equal to the sum of the products
$$\DM\sum_i C_i \Gamma_{light}^{ren} \left( O_i^{ren} \right)
\EQNO\sv{\xtwo} $$
where the $C_i$ are the Wilson coefficients, and contain all the heavy
W-boson and/or top quark effects
(also called the short distance effects) and the $O_i^{ren}$ are
local renormalized operators of dimension equal to at most six.
The subscript {\it light} is there to remind us to use
light fields ($u,d,s,c,b,G,A$) only for the internal as well as external
lines of the graphs
when computing the Green functions with insertions of operators.
In this case, $\tau^{(0)}_{\gamma _0}$ gives us the tree vertex of
Fig. 3, while $(1-\tau ^{(2)}_\Gamma )\tau^{(0)}_{\gamma _0}\Gamma $
is the one loop renormalized matrix element of this same vertex,
which is the first graph on the right hand side of Fig. 2.
$(\tau^{(2)}_\Gamma -\tau^{(0)}_\Gamma )\Gamma $ denoted by the
second graph is a part of the effective
Lagrangian induced at one loop order, where $C_i$ have been evaluated to one
loop and $\Gamma (O_i)$ have tree values. A general remark is in order.
Such evaluations of $C_i$ give us the boundary conditions for RGE.  The
matching between the high energy theory and the low energy theory
has been automatically done.

We would like to digress for a momemt and discuss the renormalization
of the Feynman graphs
in the full theory and the renormalization of the operators in the effective
theory. In Eq. (\xx) we have used momentum subtraction
for the graphs in the full
theory. In Eq. (\xfirst) we have also used momentum subtraction to renormalize
the operators: this is shown in the first term on the right in Eq. (3), where
$$\left( - \tau_{\Gamma}^{(2)} \right) \; \tau_{\gamma_{0}}^{(0)} \, \Gamma$$
is the counterterm to the bare graph $\DM \tau_{\gamma_{0}}^{(0)} \, \Gamma$.
Actually, however, the renormalization prescription of the operators in the
effective
theory may be taken to be different from that of the graphs in the full
theory. For example, we may switch to the minimal subtraction to define
the renormalized
operators. This is achieved by rewriting Eq. (\xfirst) as (neglecting
$1/m_w^4$
effects)
$$\eqalign{
\Gamma ^{(ren)} =\ \ & \left( 1 - \tau_{\Gamma}^{(\epsilon)} \right) \,
                          \,\tau_{\gamma_{0}}^{(0)} \, \Gamma  \cr
             + & \left[ \tau_{\Gamma}^{(2)} - \tau_{\Gamma}^{(0)}
             - \left( \tau_{\Gamma}^{(2)} - \tau_{\Gamma}^{(\epsilon)} \right)
		\; \tau_{\gamma_{0}}^{(0)} \right] \, \Gamma .\cr}
\EQNO\sv{\xaa} $$
What we learned from this exercise is that one can choose any scheme
to renormalize the operators. This freedom of choice is also reflected
in the dependence of the Wilson coefficients: the second term on the right
in Eq. (1) and that in Eq. (\xaa) depend on the renormalization
chosen for the corresponding first term.
However, the sum $\sum_i C_i  \Gamma_{light}^{ren}(O_i^{ren})$ is uniquely
given once the renormalization of the graphs in the full
theory is fixed: it equals $\Gamma^{ren}$. One may wonder why
we have used momentum subtraction for the full theory instead of minimal
subtraction. The reason is that power counting is easy in momentum
subtraction.  If we had used the MS scheme for the full
theory, then some of the $C_i$ would have been of order $m_w^0$, and we
would not have had {\it explicit decoupling} of the W-boson.
The effective theory obtained would have necessarily required
multiple insertions of the associated operators, because they would not
have been suppressed by inverse powers of $m_w$. In the end, all these multiple
insertions could be reabsorbed into the redefinition of the fields and
parameters in the full theory and one would obtain the same on-shell results.
On the other hand, using momentum subtraction for the heavy graphs
insures us that all the Wilson coefficients are of order $1/m_w^2$,
and therefore we have explicit decoupling. The operators are inserted
once only.  Also the derivation of renormalization group equations
is simpler when there is explicit decoupling.

\medskip
Let us now consider another example which creates "problems" in the
effective theory. This is illustrated in Fig. 4, where again we want to
work to an accuracy of $1/m_w^2$. Let $\Gamma$ be the
graph on the left in Fig. 4, and $\gamma_0$ be the tree subgraph of
$\Gamma$ which contains the $W$-boson line, similar to Fig. 3.
$\Gamma$ is finite and needs no
renormalization. The Zimmermann algebraic rearrangement gives
$$\eqalign{
\Gamma  =\ \ & \left( 1 - \tau_{\Gamma}^{(0)} \right) \,
                          \,\tau_{\gamma_{0}}^{(0)} \, \Gamma  \cr
                + & \tau_{\Gamma}^{(0)} \Gamma \cr
                + & \left( 1 - \tau_{\Gamma}^{(0)} \right) \,
                 \left(\, 1 - \tau_{\gamma_{0}}^{(0)} \right) \, \Gamma \cr}
\EQNO\sv{\xab} $$
The first and second terms on the right in Eq. (\xab) correspond
to the first and second diagrams on the right in Fig. 4 respectively.
The last term in Eq. (\xab) is of order $1/m_w^4$ and will be neglected.
Here we have again tentatively used momentum subtraction to define
the renormalized Green
functions with $O_i^{ren}$ inserted. In a moment we will rewrite Eq. (\xab)
in the MS scheme, but first we would like to make a remark.
The first and second terms on the right in Eq. (\xab) are well defined
and the Dirac algebra can be done in four dimensions safely since each term
is finite. This means that $\Gamma^{ren}(O_i^{ren})$ is unambiguously defined,
as long as we evaluate the terms in the combinations as shown.
However, we may want to compute the 'counterterms'
$\DM \left( \  - \tau_{\Gamma}^{(0)} \,
 \,\tau_{\gamma_{0}}^{(0)} \, \Gamma\ \right) $ separately by themselves, then
we encounter in dimensional regularization expressions like
$$ \eqalign{
S \DEF {1 \over \epsilon} \,\Big[ &
\left(\,\gamma^{\mu}\,\gamma^{\nu}\,\gamma^{\alpha}\,P_{\!\!{_L}}\right)_{s_1\,s_2}\;
\left(\,\gamma^{\mu}\,\gamma^{\nu}\,\gamma^{\alpha}\,P_{\!\!{_L}}\right)_{s_3\,s_4}\cr
& - 16 \left(\,\gamma^{\mu}\,P_{\!\!{_L}}\right)_{s_1\,s_2}\;
\left(\,\gamma^{\mu}\,P_{\!\!{_L}}\right)_{s_3\,s_4} \Big], \cr}
\EQNO\sv{\xac} $$
where $P_{\!\!{_{L,\,R}}} \DEF (1\pm \gamma^5)/2$.
These 'counterterms' are just the unrenormalized operator ($\tau ^{(0)}
_{\gamma _0})$ inserted matrix elements, which form the basis for
operator analysis that enters into RGE.
In four dimensions the above expression in square brackets gives zero,
but in $n\neq 4$
dimensions, it has to be defined:
is it a pole which must be subtracted, or is it
a finite quantity which should not require a subtraction if one
uses the MS scheme to renormalize the operators?
This is the main difference between the naive scheme of
Ref.[1] and the scheme of Refs.[2,3]. In Ref.[2,3], the authors
looked upon such an expression as new structures and introduced
evanescent operators for renormalizaton, while in Ref.[1] we did
not.  We shall come back to this shortly.
\medskip
Let us go back to Eq. (\xab) and rewrite it such that the operators are
renormalized in the MS scheme.
$$\eqalign{
\Gamma  =\ \
& \left( 1 - \tau_{\Gamma}^{(\epsilon)} \right) \,
                          \,\tau_{\gamma_{0}}^{(0)} \, \Gamma  \cr
                + &\left[ \tau_{\Gamma}^{(0)}
  - \left( \tau_{\Gamma}^{(0)} - \tau_{\Gamma}^{(\epsilon)} \right) \,
                          \,\tau_{\gamma_{0}}^{(0)} \,\right] \Gamma  \cr}
\EQNO\sv{\xad} $$
The $1/m_w^4$ terms have been discarded. In this scheme too, we encounter
expressions like those given in Eq. (\xac). Again, if one's intent is
just to use the Zimmermann indentities as a technique to calcualte Fig. 4,
then this is not a real problem: the expression
$\left( \tau_{\Gamma}^{(0)} - \tau_{\Gamma}^{(\epsilon)} \right) \,
                          \,\tau_{\gamma_{0}}^{(0)} \,\, \Gamma$,
which contains evanescent structures, is
found in the first and second terms on the right in Eq. (8)
with opposite signs. This makes the sum of the two terms
independent of the definition (i.e. prescription) of that evanescent
expression. One has various choices to define
the expression in Eq. (\xac), as long as one
uses the same choice to calculate the first and second
terms on the right in Eq. (\xad). Two possible prescriptions are:

\noindent
a) Formally, one can think of Eq. (\xac) as a pole
that multiplies evanescent structures, and therefore treat them as
counterterms for the operators (e.g. $\tau ^{(0)}_{\gamma _0}$) which
need renormalization. In this scheme,
one introduces evanescent operators to carry out the renormalization
of the operators, but some extra work is needed to obtain physical results.
The renormalized Green functions of these evanescent operators must be
included or be transformed away in the limit $n \rightarrow 4$.

\medskip\noindent
b) One can think of the expression in Eq. (\xac) as being finite and give a
prescription to define its
value. For instance, we can use the following definition:
$$ S = a\; \left(\,\gamma^{\mu}\,P_{\!\!{_L}}\right)_{s_1\,s_2}\;
\left(\,\gamma^{\mu}\,P_{\!\!{_L}}\right)_{s_3\,s_4}
\EQNO\sv{\xae} $$
where $a$ is arbitrary, but otherwise finite in the limit
$\epsilon \rightarrow 0$. In this scheme, there is no need
to introduce any evanescent operator. Only physical operators need
be included in the analysis. We will present in section 4
a consistent scheme to evaluate the coefficient $a$.

\medskip
Let us summarize what we have said so far. The effective theory can be
derived from the full theory with the use of the Zimmermann rearrangement:
all the heavy effects in the full theory can be extracted by
partitioning in momentum space the heavy graphs of the full theory.
This is an alternative to the operator product expansion to
construct the effective theory. In deriving the effective
theory, one encounters evanescent structures which must be
defined in the context of dimensional regularization.
This makes the renormalization of the physical
operators {\it artificially} prescription dependent, and as a consequence
the Wilson coefficients become prescription dependent as well.
However, provided
one stays being consistent in dealing with these evanescent
structures, one is garanteed that $\sum _i C_i \Gamma^{ren}_{light}(O_i^{ren})$
is independent of the prescription used to renormalize the various operators.
One is then free to either include evanescent operators
(assuming there are no inconsistencies within such a scheme)
or not include them, in which case one must be ready
to present a consistent procedure in dealing with the evanescent structures.
In section 5, we will treat in some
details the processes $b\rightarrow s$ in the LLA, and explain why
one has to be careful when evanescent operators are first introduced and
then transformed away by a finite renormalization.

\medskip
Before we close this section, let us complete this last example.  For the
process $b+\bar{c}\rightarrow \bar{c}+s$, in addition to Fig. 4 there are five
more
diagrams, plus external wave function renormalizations.  If we follow the
procedure to calculate the second term of Eq.(8) and its corresponding
ones for the other diagrams, we obtain the effective Lagrangian to
$g_s^2$

$$\eqalign { L^{eff}_{b+{\bar{c}}\rightarrow {\bar{c}}+s}=&-\,G_c \
\bar c \,\gamma _\mu P_{\!\!{_L}} \,s \ \
\bar s \,\gamma ^\mu P_{\!\!{_L}} \,b \cr
& + \, G_c \; {g_s^2\over 16\pi^2} \;
\Big[\;
\bar c \,\gamma _\mu P_{\!\!{_L}} {\lambda_a \over 2} \,b \ \
\bar s \,\gamma ^\mu P_{\!\!{_L}} {\lambda_a \over 2} \,c \ \
\left(\, 6 \,\log {m_w^2\over \mu^2} \,-\, 11 \,\right)\cr
& -\ \bar c \,R_{\kappa \lambda }\gamma _\mu P_{\!\!{_L}}
{\lambda _a\over 2}\,b \ \
\bar s \, \gamma ^\mu R^{\lambda \kappa }P_{\!\!{_L}}
{\lambda _a \over 2}\, c \
\left(\,{1\over 2}\,\log {m_w^2\over \mu^2} \,-\,{1\over 4}\,\right)
\;\Big],\cr}
\EQNO\sv{\xaf}$$
in which the evanescent operator is
$$\eqalign {R_{\kappa \lambda}\gamma_\mu P_{\!\!{_L}}
\otimes \gamma^\mu R^{\lambda \kappa }P_{\!\!{_L}} &\equiv
\gamma _\kappa \gamma _\lambda \gamma _\mu P_{\!\!{_L}} \otimes
\gamma ^\mu \gamma^ \lambda \gamma ^\kappa P_{\!\!{_L}}
-\gamma _\kappa \gamma _\lambda \gamma _\mu P_{\!\!{_L}}
\gamma^ \lambda \gamma ^\kappa \otimes \gamma ^\mu  P_{\!\!{_L}} \cr &
-\gamma ^\mu  P_{\!\!{_L}}\otimes\gamma _\kappa \gamma _\lambda
\gamma _\mu P_{\!\!{_L}}
\gamma^ \lambda \gamma ^\kappa
+\gamma _\mu \gamma _\lambda \gamma _\kappa P_{\!\!{_L}} \otimes
\gamma ^\kappa \gamma^ \lambda \gamma ^\mu P_{\!\!{_L}}, \cr} \EQNO\sv{\xag}$$
$G_c={g_w^2\over 2m_w^2}V_{cs}^{\star }V_{cb},$ and the V's are
the relevant Cabbibo-Kobayashi-Maskawa matrix elements.  This agrees with the
result given by Buras and Weisz in Ref.[7]. The constant -11 in the first term
of Eq.({\xaf}) is the one loop matching condition for the physical
Wilson coefficient and it
depends on the choice of the evanescent operator of Eq.({\xag}).  It is needed
for subleading logarithm approximation in solving RGE.
Please note that the first
term of Eq. (8) is also dependent on evanescent operators.  It is
only the S-matrix elements which should be and will be basis independent.

We have also explicitly checked that the method is internally
consistent up to two loop order in Feynman diagram evaluation of the
processes $b\rightarrow s \gamma$ and $b \rightarrow s G.$ The results
can be found in Ref. [4].

\bigskip \noindent
{\bf 3. Renormalization Group Equations (RGE)}
\bigskip

We want to give a quick review in this section, so that we can establish
notation.  We will define the renormalized operators and their anomalous
dimension matrix. Then we will write down the RGE for the Wilson
coefficients and that for the renormalized
Green functions with the insertion of an operator. We will also discuss
the finite transformations that take us from one renormalization scheme
to another, and the advantage of using a {\it mass independent}
renormalization scheme (MIRS) for the operators. The meaning of the leading
logarithmic approximation (LLA) and the next-to-leading logarithmic
approximation (NLLA) will be briefly reviewed in such a scheme. We will
also say a few words about the matching of the effective theory
with the full theory.  It is understood that we use dimensional
continuation as regularization.

\medskip
The renormalized operators are defined as linear combinations of a
complete set of bare operators
$$O_{\ii}^{ren} \DEF Z_{\ii\jj} O_{\jj}^{bare}  \EQNO\sv{\xO3}  $$
where the matrix $Z$ is an infinite square matrix if we introduce
evanescent operators, otherwise it is finite. It is determined order
by order in perturbation theory by requiring
that $\Gamma^{ren}(O_{\ii}^{ren})$ be finite in the
limit $n\rightarrow 4$. We would like to mention
that the infinite matrices we will encounter have entries which
eventually are unbounded (this will be  discussed in section 5); we can
therefore treat these matrices very formally.

\medskip
The $\mu$-independence of bare quantities
allows us to derive a RGE for the Green functions with the insertion
of a renormalized operator:
$$\eqalign {
& \mu {d \over d \mu}\Gamma^{bare}(O_{\ii}^{bare}) =
0 \ \ \ \Longrightarrow \cr
& \left[\left( \mu {d \over d \mu} + N_{\phi} \gamma_{\phi} \right)
 \delta_{\ii\jj} + \gamma_{\ii\jj} \right] \Gamma^{ren}(O_{\jj}^{ren}) = 0 \cr
} \EQNO\sv{\xRGE}  $$
where $N_{\phi}\gamma_{\phi}$ denotes symbolically the anomalous
dimension of the external fields, and $\gamma_{\ii\jj}$ is the
anomalous dimension mixing matrix of the operators given by
$$ \gamma_{\ii\jj} = - \left( \mu {d \over d \mu} Z_{\ii k} \right)
Z^{-1}_{k\jj}. \EQNO\sv{\xgamma}  $$
The same mixing matrix also governs the evolution of the Wilson
coefficients $C_{\ii}$
$$\mu {d C_{\ii}\over d \mu} = C_{\jj} \gamma_{\jj\ii}, \EQNO\sv{\eqCI} $$
which is a result of the $\mu$-independence of $C_iO_i$.
Eqs. ({\xO3}-{\eqCI}) are valid for all schemes,
including {\it mass dependent}
renormalization schemes. They are also valid (at least formally)
if we include evanescent
operators in the analysis.

\medskip

A change in the renormalization of the operators
must be compensated by a change in the Wilson coefficients. This amounts
to making a finite transformation such that the sum of products
$ C_i O_i^{ren}$
is invariant.  We shall drop the upperscript ${ren}$ from now on.  Under
a finite renormalization (1+A), various quantities are transformed into the
corresponding primed ones
$$ \eqalign{\baselineskip=18pt
& O' = (1+A) \, O = (1+A) \, Z O^{(bare)} \DEF Z' O^{(bare)}, \cr
& C' = C \, (1+A)^{-1}, \cr
& \gamma ' \DEF - \mu {d Z'\over d \mu} Z'^{-1} =
 - \left[ \mu {d  \over d\mu} (1+A) \right]\; (1+A)^{-1}
   + (1+A)\, \gamma \, (1+A)^{-1},  \cr
} \EQNO\sv{\xth35} $$
where the matrix $A$ is assumed finite but otherwise arbitrary, and may depend
explicitly on $\mu$.

\medskip

We will now discuss the advantages of using
a mass independent scheme. In such a scheme, the Wilson coefficient vector
$C$ is $\underline{\rm assumed}$ to have the following expansion:
$$ C = \sum_{n=0}^{\infty}
   \left( g^2 \log {m_{h}^2 \over \mu^2 } \right)^n
   \left[ c^{(0)}_n + c^{(1)}_n g^2 + \dots + c^{(p)}_n  g^{2p} + \dots
   \right]  \EQNO\sv{\formC}  $$
where $m_{h}$ stands for the mass of the heavy field which is
integrated out, and $g \DEF g(\mu)$.
In Eq. $(\formC)$ we have shown the explicit dependence of $C$ on $g$,
$m_h$ and $\mu$; the $c^{(p)}_n$ do not depend explicitly on $g$,
$m_h$ and $\mu$.
In the LLA we keep $c^{(0)}$ only, in the NLLA we keep both $c^{(0)}$
and $c^{(1)}$, etc ...
$$\eqalign{\baselineskip=18pt
& C^{^{LLA}} = \sum_{n=0}^{\infty} c^{(0)}_n
   \left( g^2 \log {m_{h}^2 \over \mu^2 } \right)^n, \cr
& C^{^{NLLA}} = \sum_{n=0}^{\infty} \left( c^{(0)}_n + g^2 c^{(1)}_n \right)\
   \left( g^2 \log {m_{h}^2 \over \mu^2 } \right)^n, \cr
& \dots \cr } \EQNO\sv{\eqtion}  $$
To solve the RGE of Eq. ({\eqCI}) for $C$ we need the initial (or boundary)
conditions
$C(\mu)|_{\mu=m_h}$. In a mass independent scheme, the initial conditions
are given by $C(m_h)=c^{(0)}_0$ for the LLA,
and $C(m_h)=c^{(0)}_0 + g(m_h)^2 c^{(1)}_0$ for the NLLA.  These are all
very simple and clean in principle.  As a contradistinction,
in a mass dependent scheme where the $c^{(p)}$ have an explicit
dependence on $\mu$, the boundary conditions become $C(m_h)=
\sum_{p=0}^{\infty} g(m_h)^{2p} \, c^{(p)}_0(m_h)$, and one cannot truncate
the series, say to the first term to obtain the LLA result.  This is because
each term $g(m_h)^{2p}\,c^{(p)}_0(m_h)$ contains
``$\left[g^2 \log(m_h^2)\right]^n$'' and therefore has to be taken into
account to fulfill a consistent partial sum for the LLA.

\medskip

As we saw, in a mass-independent scheme for the NLLA,
we need to compute
 $c^{(1)}_0$ as well to obtain the
boundary conditions. We would like to repeat that these NLLA initial
conditions ($c^{(1)}_0$) are scheme-dependent, i.e. they depend on the
particular choice of a mass
independent scheme one uses to renormalize the operators. Therefore
the $C(\mu)$ we obtain in the NLLA are scheme-dependent as well.
Again, the sum $C_i(\mu) O_i(\mu)$ is
scheme-independent.  For instance,
for four fermion processes, the NLLA results are given by:
$$ \left[ C_i \Gamma(O_i) \right]^{^{NLLA}} =
   C_i^{^{NLLA}} \Gamma(O_i)^{(tree)} + C_i^{^{LLA}}
\Gamma(O_i)^{(1-loop)}.
\EQNO\sv{\xnlla}  $$
This is illustrated in Fig. 5.  In this regard, we differ from a remark
in Ref.[8] by Dugan and Grinstein.
In Eq. (\xnlla) the scheme-dependence of the $C^{^{NLLA}}$ will be
precisely canceled by that of $\Gamma(O_i)^{(1-loop)}$.
We have already given an explicit example at the end of the last
section in this regard.
\medskip
In the renormalization group analysis of $b\rightarrow s$ processes,
all authors have assumed that their schemes are mass-independent.  A
closer examination seems warranted, however.

\bigskip
\noindent
{\bf 4. Evanescent structures in the t'Hooft-Veltman scheme}
\medskip

In this section, we present in the t'Hooft-Veltman scheme
certain equations which are needed to define the renormalized graphs with an
arbitrary number of external fermions, with or without insertions of
operators. We are particularly interested in defining the expression
in Eq. (\xac) without introducing any evanescent operator.

\smallskip
In the t'Hooft-Veltman scheme, the external particles are kept in four
dimensions: their momenta have the form
$\DM p_{ext} = (p_1,p_2,p_3,p_4,0,\dots,0)$, and the Lorentz indices
attached to them run from 1 to 4. The spinors
attached to these external particles will be elaborated.
We will first solve the Dirac equation in even integer $n$ dimensions.
Among the solutions we obtain,
we will {\it choose} two to describe
the two physical degrees of freedom of an external particle
and two others for the anti-particle.
These particular solutions will have the property that only the first
four components of are non-zero. This will allow us to
write for the spinors of the external particles as
$$u(p)={\cal P}\,u(p) \ \ \ \ , \ \ \ \ \bar{u}(p)=\bar{u}(p){\cal P},
\EQNO\sv{\eqtion} $$
where $\cal P$ is some projector which will be defined in a moment.
This projector, which was first introduced in Ref [9],
will be used to define expressions like
those given in Eq. (\xac).

\medskip
We start with $n$ hermitian matrices  $\gamma^{\mu}$ ($\mu=1,2,\dots,n=$even),
satisfying
$$\{\gamma^{\mu},\gamma^{\nu} \}\ = \ 2\,\delta_{\mu \nu}. \EQNO\sv{\xanti} $$
These are $ 2^{n/2} \times 2^{n/2} $ matrices.
Their construction is inductive and is explained in Ref. [10].
Let us briefly review the main ideas.
The induction starts at $m=2$ (four dimensional Dirac matrices).
We then assume that we have constructed $2m$ hermitian
matrices $\gamma^{\mu}_{(2m)}$, of dimension $2^{m} \times 2^{m}$,
which satisfy Eq. (\xanti).
This is extended to  $2m+2$ hermitian matrices
$\gamma^{\mu}_{(2m+2)}$, of dimension $2^{m+1} \times 2^{m+1}$, which
satisfy Eq. (\xanti) as follows. We choose
$$ \gamma^{\mu}_{(2m+2)} = \pmatrix{
\gamma^{\mu}_{(2m)}	& 0			& \cr
0			& \gamma^{\mu}_{(2m)}	& \cr}
\ \ \ \ {\rm if}\ \ \ \ \ 1 \le \mu \le 2m, $$
and
$$ \gamma^{2m+1}_{(2m+2)} = \pmatrix{
0			& \hat{\gamma}_{(2m)}	& \cr
\hat{\gamma}_{(2m)}	& 0			& \cr}\;,
\hskip1truecm
\gamma^{2m+2}_{(2m+2)} = \pmatrix{
0			& i\hat{\gamma}_{(2m)}	& \cr
-i\hat{\gamma}_{(2m)}	& 0			& \cr}, $$
where
$$ \hat{\gamma}_{(2m)} =\,-\,i^{m} \,
\gamma^1_{(2m)} \dots \gamma^{2m}_{(2m)}. $$
One can easily check that all the matrices $\gamma^{\mu}_{(2m+2)}$
are hermitian, and satisfy Eq. (\xanti).

{}From now on, we will write $\gamma^{\mu}$ instead of $\gamma^{\mu}_{(n)}$.
In this representation, the first four matrices
$\gamma^i$, $i=1,\dots,4$, have the special form:
$$ \gamma^i = \pmatrix{
             \gamma^i_{(4)}  &          &            & \cr
                       & \gamma^i_{(4)} &            & \cr
                       &          & \ddots     & \cr }, \EQNO\sv{\eqtion} $$

\noindent
where $\gamma^i_{(4)}$ are the $ 4 \times 4 $ Dirac $\gamma$-matrices,
chosen such that $\gamma^4_{(4)}$ is diagonal:
$$ \gamma^4_{(4)} = \pmatrix{
  1 &    &   &  \cr
    & 1  &   &  \cr
    &    &-1 &  \cr
    &    &   & -1 \cr}.
\EQNO\sv{\xgam4} $$
The non-trivial solutions of the Dirac equation, i. e. those which
do not vanish identically, which satisfy
$$ \eqalign{
& \left( i\gamma . p + m \right) u(p) = 0,  \cr
& p^2 + m^2 = 0, \cr }\EQNO\sv{\xdirac} $$
are easily found to be
$$\cases{
 u^{(i)}(p) \sim \left( -i\,\gamma . p + m \right ) V^{(i)}, \cr
\left[V^{(i)}\right]_j \DEF \delta_{ij} \hskip1truecm ;
\hskip1truecm i,j=1,2,3,4,\dots,2^{n/2}. \cr}\EQNO\sv{\xspinP} $$

Those with $p_4 = + i E, \ \ E=\sqrt{\vec{p}\,^2+m^2}$, will correspond
to particles, while others
with $p_4=-i E$ will be for anti-particles.
For concreteness we will discuss the particle-solutions only,
where the $2^{n/2-1}$
independent spinors have been afixed with a normalizing factor,
$$ \cases{
\DM u^{(i)}(p) = { 1\over \sqrt{ 2 E \left( m+E \right) } }
        \left( -i\gamma . p + m \right) V^{(i)}, \cr
\DM i=1,2,5,6,9,10,\dots,4\ell +1,4\ell +2,\dots \cr}\EQNO\sv{\xspinP} $$
$0 \le \ell \le 2^{n/2-2}-1$.
The normalization of the spinors is:
$$ \left( u ^{(i)} \right)^{\dagger} \; u^{(i)} = 1,
\ \ \ \ {\rm for\ each}\ i \EQNO\sv{\eqtion}  $$
and one can easily derive:
$$ \eqalign{
& \overline{u}^{(i)}  \; u^{(i)} = {m \over E},
   \hskip0.8truecm {\rm for\ each\ } i, \cr
& \sum_i u^{(i)} \, \overline{u}^{(i)} = { - i \gamma . p + m \over 2 E},\cr}
\EQNO\sv{\eqtion} $$
where $i$ takes the values $1,2,5,6,\dots,4\ell +1,4\ell +2,\dots$.
In deriving the above equation, we have used
$$ \sum_{i} V^{(i)}\,V^{(i)\,t} = {1 +\gamma^4 \over 2}, $$
where the upperscript $^t$ stands for transpose.

\medskip
If we go to the rest frame of the particle ($\vec{p}=0$ and $p^4 = + im$),
the spinors in Eq. (\xspinP) become:
$$\eqalign{
& u^{(1)} = (1,0,0,0,0,0,0,0,0,\dots,0),\cr
& u^{(2)} = (0,1,0,0,0,0,0,0,0,\dots,0),\cr
& u^{(5)} = (0,0,0,0,1,0,0,0,0,\dots,0),\cr
& u^{(6)} = (0,0,0,0,0,1,0,0,0,\dots,0),\cr
& \dots \cr} \EQNO\sv{\xspinR}  $$
In the real world where $n=4$, a spin-${1\over 2}$ particle has only
two degrees of freedom.  This nomenclature is kept
in $n$ dimensions where we {\it choose} to designate the spinors
$u^{(1)}$ and $u^{(2)}$ given in Eq. (\xspinR) as the
two $\underline{\rm physical}$ degrees of freedom of a particle at rest:
$u^{(1)}$ for a particle with spin up, and
$u^{(2)}$ with spin down.
The other spinors will describe non-physical degrees of freedom
which must disappear when the limit $n\rightarrow 4$ is taken.
Similarly, we will choose $u^{(3)}(p)$ and $u^{(4)}(p)$,
$\vec{p}=0,\ p_4=-im$, to describe the
anti-particle at rest, with spin up and down, respectively.

Therefore, when computing Feynman diagrams in $n$ dimensions,
because the external fermions are kept in four dimensions,
they will be described by states
(or spinors) which are linear combinations of
$u^{(1)}(p)$ and $u^{(2)}(p)$ .
Such states, which we will call {\it physical spinors}, have the form:
$$ u(p)_{phys}= \alpha \; u^{(1)}(p) + \beta \; u^{(2)}(p) =
({\rm x},{\rm x},{\rm x},{\rm x},0,\dots,0) \EQNO\sv{\xphysS} $$
$i.e.$ only the first four components of $u_{phys}$ can be non-zero.
This can easily be checked if we note that when $p$ is the momentum
of a particle in four dimensions, then the matrix $\gamma . p$ is of the form
$$ \gamma . p = \sum_{\mu=1}^{4} \gamma^{\mu} p^{\mu} = \pmatrix{
             \gamma_{(4)} . p  &          &            & \cr
                       & \gamma_{(4)} . p &            & \cr
                       &          & \ddots     & \cr } \EQNO\sv{\xD4} $$
and therefore the vectors $(-i\gamma . p + m)V^{1,2}$ are of the form
shown in Eq. (\xphysS).
We now define a $2^{n/2} \times 2^{n/2}$
matrix ${\cal P}$
$$ {\cal P} \DEF \pmatrix{
	I_{4}	&  0 \cr
		0	&  0 \cr}. \EQNO\sv{\eqtion} $$
In the above equation $I_{4}$ is the $4 \times 4$ identity matrix.
${\cal P}$ is a projector (${\cal P}{\cal P} = {\cal P}$) which satisfies:
$$u(p)_{phys}={\cal P}\;u(p)_{phys} \ \ \ \ ,
\ \ \ \ \bar{u}(p)_{phys}=\bar{u}(p)_{phys}\;{\cal P}.
\EQNO\sv{\eqtion} $$

We can now state our prescription in handling gamma matrices:
''when computing Feynman diagrams
in the t'Hooft-Veltman scheme, we simply attach a factor ${\cal P}$ to
each external fermion'' (see Eq. (32)). For instance, for four quark
processes, Eq. (\xac) will be rewritten as:
$$ \eqalign{
S = {1 \over \epsilon} \,\Big[ &
\left(\,{\cal P}\;\gamma^{\mu}\,\gamma^{\nu}\,\gamma^{\alpha}\,
P_{\!\!{_L}}\;{\cal P}\right)
_{s_1\,s_2}\;
\left(\,{\cal P}\;\gamma^{\mu}\,\gamma^{\nu}\,\gamma^{\alpha}\,
P_{\!\!{_L}}\;{\cal P}\right)
_{s_3\,s_4}\; \cr
& - 16
\left(\,{\cal P}\;\gamma^{\mu}\,P_{\!\!{_L}}\;{\cal P}\right)_{s_1\,s_2}\;
\left(\,{\cal P}\;\gamma^{\mu}\,P_{\!\!{_L}}\;{\cal P}\right)_{s_3\,s_4}
\Big]. \cr}
\EQNO\sv{\eqtion} $$
We will see later that the presence of ${\cal P}$ will make the above
expression collapse into Eq. (9), in fact with a=0.  In other words,
the gamma algebra in the first term may as well have been done in
four dimensions.

\medskip
One can check that for $n$ even, ${\cal P}$ can be written as:
$$ \eqalign{
{\cal P} &  = \;
\left( { 1-i\,\gamma^{\Eb}_5 \gamma^{\Eb}_6 \over 2 }  \right) \;
\left( { 1-i\,\gamma^{\Eb}_7 \gamma^{\Eb}_8 \over 2 }  \right) \;
\dots
\left( { 1-i\,\gamma^{\Eb}_{n-1} \gamma^{\Eb}_{n} \over 2 }  \right) \cr
    & = \; \prod_{p=3}^{n/2} \;
\left( { 1-i\,\gamma^{\Eb}_{2p-1} \gamma^{\Eb}_{2p} \over 2 }  \right). \cr}
\EQNO\sv{\defPg} $$
This is easily seen if we observe that the products of matrices
$\ \gamma_{2p-1}\gamma_{2p}$, $p\ge 3$ are diagonal; if $I_m$ and $0_m$ denote
identity and the null $2^m\times 2^m$ matrices, respectively, we have:
$$\eqalign{
& {1-i\gamma_{n-1}\gamma_{n} \over 2} = \pmatrix{
	I_{2^{n/2-1}}	&   \cr
			& 0_{2^{n/2-1}} \cr}, \cr
& {1-i\gamma_{n-3}\gamma_{n-2}\over 2} = \pmatrix{
	I_{2^{n/2-2}}	&  & & \cr
	& 0_{2^{n/2-2}} & & \cr
	& & I_{2^{n/2-2}} & \cr
	& & & 0_{2^{n/2-2}} \cr }, \cr
& \ldots \cr} \EQNO\sv{\xthreetwo} $$
Using  Eq. (\defPg), we easily derive:
$$\eqalign{
& \left[{\cal P},\gamma^{\mu} \right]=0 \hskip1truecm
{\rm if\ \ } \mu=1,2,3,4, \cr
& {\cal P}\,\gamma^{\mu}\,{\cal P} = 0  \hskip1truecm {\rm if\ \ } \mu > 4, \cr
& \left[ {\cal P}, \gamma^{\,\underline{5} } \right] = 0. \cr}
\EQNO\sv{\comrel} $$
The last identity in Eq. (\comrel) holds if $\gamma^{\,\underline{5}}$
anticommutes with all $\gamma^{\mu}$. It also holds in the t'Hooft-
Veltman definition, where
$\gamma^{\,\underline{5}}$ anticommutes with the first four $\gamma$-matrices
but commutes with the remaining $n-4$ matrices. This matrix should not be
confused with the matrix $\gamma^{\mu}$, when $\mu=5$.
Moreover, whether $\gamma^{\,\underline{5}}$ is defined as the product
of the first four $\gamma$-matrices, or as the product of the
$n$ $\gamma$-matrices ($n$ even), we have:
$${\cal P}\,\gamma^{\underline{5}} \, {\cal P} =
{\cal P} \, \gamma^1 \gamma^2 \gamma^3 \gamma^4 \, {\cal P} =
{\epsilon^{\,\bar{\mu}_1 \, \bar{\mu}_2 \, \bar{\mu}_3 \,
 \bar{\mu}_4 } \over 4 !} \;
{\cal P}\, \gamma^{\bar{\mu}_1} \gamma^{\bar{\mu}_2}
\gamma^{\bar{\mu}_3} \gamma^{\bar{\mu}_4} \, {\cal P}, \EQNO\sv{\eqtion} $$
where the barred indices take on the values 1,2,3 and 4 only.

\medskip
As far as four fermion processes are concerned, we need to evaluate
the following quantities:
$${\cal{T}}_{ij} \DEF
\left( {\cal P} {\cal S}_i {\cal P} \right)_{s_1\, s_2}\
\left( {\cal P} {\cal S}_j {\cal P} \right)_{s_3\, s_4}\
\EQNO\sv{\xthreethree} $$
where the ${\cal S}_i$ are products of $\gamma$-matrices, and possibly
$\gamma^{\underline {5}}$'s. The algorithm to compute such expressions
is the following: we first use the sum splitting notation
$$ \left( \gamma^{\mu} \right)_{s_1 \,s_2}\;
   \left( \gamma^{\mu} \right)_{s_3 \,s_4}\ = \
   \left( \gamma^{\bar{\mu}} \right)_{s_1 \,s_2}\;
   \left( \gamma^{\bar{\mu}} \right)_{s_1 \,s_2}\ + \
   \left( \gamma^{\hat{\mu}} \right)_{s_1 \,s_2}\;
   \left( \gamma^{\hat{\mu}} \right)_{s_1 \,s_2}, \EQNO\sv{\eqtion} $$
where
$$ \eqalign{
  & \mu = 1,\dots,n,  \cr
  & \bar{\mu} = 1,\dots,4, \cr
  & \hat{\mu} = 5,\dots,n. \cr} \EQNO\sv{\xindices}  $$
Then we move all $\gamma^{\bar{\mu}_i}$,
to the left with the use of Eq. ({\xanti}).
If $\gamma^{\underline{5}}$'s with the t'Hooft-Veltman prescription
are present, they are also moved
to the left. [For that matter, one could use a completely
anti-commuting $\gamma^{\underline{5}}$ if there are no traces involving
$\gamma^{\underline{5}}$]. At this point, ${\cal{T}}_{ij}$ has the form
$$ \eqalign{ {\cal{T}}_{ij} \sim &
\left[\;
\gamma^{\underline{5}} \dots \gamma^{\underline{5}}\,
\gamma^{\bar{\mu}_1} \gamma^{\bar{\mu}_2}\dots \ \gamma^{\bar{\mu}_m}
\left( {\cal P} \gamma^{\hat{\nu}_1}\gamma^{\hat{\nu}_2} \dots
\gamma^{\hat{\nu}_p}{\cal P}  \right) \right]_{s_1\,s_2}  \cr
& \hskip1truecm \left[\;
\gamma^{\underline{5}} \dots\gamma^{\underline{5}}\,
\gamma^{\bar{\alpha}_1} \gamma^{\bar{\alpha}_2}\dots \ \gamma^{\bar{\alpha}_l}
\left( {\cal P} \gamma^{\hat{\nu}_1}\gamma^{\hat{\nu}_2} \dots
\gamma^{\hat{\nu}_p}{\cal P}
     \right) \right]_{s_3\,s_4}. \cr } \EQNO\sv{\xtij}  $$
In the expression above, we do not have any hatted index which is not
summed, because all the external particles, and hence free
indices, are in four dimensions.
The following identity is then used to reduce Eq. (\xtij):
$$\eqalign{
& \left(
{\cal P}\,\gamma^{[\hat{\nu}_1}\gamma^{\hat{\nu}_2}
\dots \gamma^{\hat{\nu}_m ]}
\,{\cal P} \right)_{s_1 \, s_2} \;
\left(
{\cal P}\,\gamma^{[\hat{\nu}_1} \gamma^{\hat{\nu}_2}
\dots \gamma^{\hat{\nu}_m ]}
\,{\cal P} \right)_{s_3 \, s_4} \;=  \cr
& \hskip2truecm \cases{ \baselineskip=18pt
  \DM ({\cal P})_{s_1\,s_2} \, ({\cal P})_{s_3\,s_4} \
 {(-1)^{m/2}\, m!\,(n/2-2)! \over
  (n/2-2-m/2)!\,(m/2)! } & \hskip0.5truecm if $m$ even,\cr
 0 & \hskip0.5truecm if $m$ odd. \cr} \cr} \EQNO \sv{\mainI} $$
where the square brackets denote antisymmetrization in all $m$ enclosed indices
(including the usual $1/m!$ factor).
At this stage, the remaining Dirac $\gamma$-algebra is performed in four
dimensions to obtain ${\cal T}_{ij}$.

Let us be reminded that Eq. (\mainI) has been
obtained for even integer, and is used as
a definition for complex values of $n$. It is easily obtained
if we observe that:
$$ {\cal P} \;\left(-i\gamma_{2p-1}\gamma_{2p} \right)\;{\cal P} \,=\,
{\cal P} \hskip1.4truecm {\rm for\ \ } p \ge 3 \EQNO $$
As an example, an application of Eq. (\mainI) gives
$$ \left(
{\cal P}\,\gamma^{\hat{\nu1}_1} \gamma^{\hat{\nu}_2} \,{\cal P}
\right)_{s_1 \, s_2} \;
\left(
{\cal P}\,\gamma^{\hat{\nu}_1} \gamma^{\hat{\nu}_2} \,{\cal P}
\right)_{s_3 \, s_4} \ =\ 0 \EQNO \sv{\xthreeeight} $$
where there is no antisymmetrization over $\hat{\nu}_1$ and $\hat{\nu}_2$.
The final result for ${\cal T}_{ij}$
must then have the form:
$$ {\cal T}_{ij} \,\sim \, \sum_{A,B=1}^5 t_{\!_{A\,B}} \;
\left( \Gamma_{\!_A} \right)_{s_1\,s_2} \
\left( \Gamma_{\!_B} \right)_{s_3\,s_4}, \EQNO \sv{\finalT}$$
where
$$\Gamma_{_A} \ \DEF \ {\cal P},\ \; {\cal P}\gamma^{\bar{\mu}} {\cal P},\ \;
{\cal P} \sigma^{\bar{\mu} \bar{\nu} } {\cal P},\ \;
{\cal P} \gamma^{\bar{\mu}} \gamma^{\underline{5}} {\cal P},\ \;
{\cal P} \gamma^{\underline{5}} {\cal P}\ \EQNO\sv{\eqtion} $$
This is obvious because any $\gamma$-string ${\cal P} {\cal S} {\cal P}$
can be written as
$$ {\cal P} {\cal S} {\cal P} = \pmatrix{
{\cal S}_4	& 0			& \cr
0			& 0	& \cr}, \EQNO $$
where $S_4$ is the projected $4\times 4$ submatrix.  Thereupon, it can be
written as a linear combination of the $\Gamma _A$ listed above.  One can
also perform ''four dimensional Fierz transformation'' for the
$\Gamma_{\!_A}$'s if that is necessary.

\medskip
Before we close this section, we would like to make a few more remarks.

\medskip\noindent
1. The final result for ${\cal T}$ is not fully Lorentz invariant (although
it is invariant under a ''four dimensional'' Lorentz transformation).
One could however rewrite Eq. (\finalT) with the use the following relations:
$$ \eqalign{
& {\cal P} \gamma^{\bar{\mu}}  {\cal P} \otimes
{\cal P}\gamma^{\bar{\mu}} {\cal P} \;=\;
{\cal P} \gamma^{\mu}  {\cal P} \otimes
{\cal P}\gamma^{\mu} {\cal P}\cr
& {\cal P} \sigma^{\bar{\mu}\bar{\nu}} {\cal P} \otimes
{\cal P} \sigma^{\bar{\mu}\bar{\nu}} {\cal P} \ =\
 {\cal P} \sigma^{\mu\nu} {\cal P} \otimes
{\cal P} \sigma^{\mu\nu}{\cal P} +
a_0 \; {\cal P} \otimes {\cal P} \cr} \EQNO \sv{\fourdim}$$
where $a_0$ is proportional to $\epsilon$. This apparently ''restores''
the full Lorentz invariance (apart from the projector ${\cal P}$ which
can then be reabsorbed into the external spinors: ${\cal P}\,u(p)=u(p)$).

\medskip\noindent
2. The spinors $u^{(1)}(p)$ and $u^{(2)}(p)$ which describe
external particles, satisfy:
$$ \eqalign{
\sum_{i=1}^2 u^{(i)}(p) \, \bar{u}^{\;(i)}(p) &=
 {1\over 2\,E\,(m+E)}
\left( -i\gamma . p + m \right)
\;{\cal P} \; {1+\gamma^{\Eb}_4 \over 2} \;{\cal P}\;
\left( -i\gamma . p + m \right) \cr
&= \;{\cal P} \; {1\over 2\,E} \left( -i\gamma . p + m \right) \;{\cal P}\cr}
\EQNO \sv{\phases}$$
The above relation is easily obtained if we observe
that $p=(p_1,p_2,p_3,p_4,0,\dots,0)$,  and
$$ \sum_{i=1}^2 V^{(i)} \, V^{(i)\,t} \ =\ {\cal P} {1+\gamma_4 \over 2}
{\cal P} \EQNO $$
Eq. (\phases) is used to calculate cross-sections, as an example.

\medskip\noindent
3. If we have more than four external fermions, we have to
generalize Eq. (\mainI). For example, expressions like
$$ \left(
{\cal P}\,\gamma^{[\hat{\mu}_1}
\dots \gamma^{\hat{\mu}_{2m} ]}
\,{\cal P} \right) \otimes
\left(
{\cal P}\,\gamma^{[\hat{\nu}_1}
\dots \gamma^{\hat{\nu}_{2n} ]}
\,{\cal P} \right) \otimes
\left(
{\cal P}\,\gamma^{[\hat{\alpha}_1}
\dots \gamma^{\hat{\nu}_{2m} ]}
\,{\cal P} \right) \EQNO $$
are first calculated in even integer $n$ dimensions. The result is then
taken as the definition of that expression for complex values of $n$.
The Lorentz indices in the above equation must of course come in pair.
Here are a few examples:
$$\eqalign{
& \left(
{\cal P}\,\gamma^{[\hat{\mu}_1}
 \gamma^{\hat{\mu}_{2} }
 \gamma^{\hat{\mu}_{3} }
 \gamma^{\hat{\mu}_{4} ]}
\,{\cal P} \right) \otimes
\left(
{\cal P}\,\gamma^{[\hat{\nu}_1}
 \gamma^{\hat{\nu}_{2} }
 \gamma^{\hat{\mu}_{3} }
 \gamma^{\hat{\mu}_{4} ]}
\,{\cal P} \right) \otimes
\left(
{\cal P}\,\gamma^{[\hat{\mu}_1}
 \gamma^{\hat{\mu}_{2} }
 \gamma^{\hat{\nu}_{1} }
 \gamma^{\hat{\nu}_{2} ]}
\,{\cal P} \right) = \cr
&\hfill -3(n-4)(n-6)(n-8)\ {\cal P} \otimes {\cal P} \otimes {\cal P},\cr
& \left(
{\cal P}\,\gamma^{[\hat{\mu}_1}
 \gamma^{\hat{\mu}_{2} }
 \gamma^{\hat{\mu}_{3} }
 \gamma^{\hat{\mu}_{4} ]}
\,{\cal P} \right) \otimes
\left(
{\cal P}\,\gamma^{[\hat{\mu}_1}
 \gamma^{\hat{\mu}_{2} ]}
\,{\cal P} \right) \otimes
\left(
{\cal P}\,\gamma^{[\hat{\mu}_3}
 \gamma^{\hat{\nu}_{4} ]}
\,{\cal P} \right) = \cr
&\hfill (n-4)(n-6)\ {\cal P} \otimes {\cal P} \otimes {\cal P},\cr
& \left(
{\cal P}\,\gamma^{[\hat{\mu}_1}
 \gamma^{\hat{\mu}_{2} ]}
\,{\cal P} \right) \otimes
\left(
{\cal P}\,\gamma^{[\hat{\mu}_2}
 \gamma^{\hat{\mu}_{3} ]}
\,{\cal P} \right) \otimes
\left(
{\cal P}\,\gamma^{[\hat{\mu}_3}
 \gamma^{\hat{\mu}_{1} ]}
\,{\cal P} \right) = 0. \cr} \EQNO $$

\medskip\noindent
4. We now give an example at the two-loop order to show how
the projector ${\cal{P}}$ can be used to derive the effective theory
to this order. This is illustrated in Fig. 6. Let $\Gamma$
represent the two-loop diagram on the left , $\gamma_0$
its tree (heavy) subgraph, and $\gamma_1$ its one-loop (heavy) subgraph.
$\gamma_0$ is represented by the diagram on the left in Fig. 3,
and $\gamma_1$ by the diagram on the left in Fig. 4.
Since $\Gamma$ is finite, the Dirac $\gamma$-algebra can be
done in four dimensions with no ambiguities. We can also attach
a factor ${\cal P}$ to each external fermion without changing
the value of the graph in the limit $n\rightarrow 4$.
The Zimmermann algebraic rearrangement for $\Gamma$ gives:
$$ \eqalign{
\Gamma =&\
\left[ \left(1-\tau_{\Gamma}^{(0)} \right)
\left(1-\tau_{\gamma_1}^{(0)} \right)
\tau_{\gamma_0}^{(0)} \right]\ \Gamma \cr
&+\
\left[ \left(1-\tau_{\Gamma}^{(0)} \right)
\tau_{\gamma_1}^{(0)} \right] \Gamma \cr
&+\ \left[ \tau_{\Gamma}^{(0)} \right] \Gamma \cr
&+\ \left[
\left(1-\tau_{\Gamma}^{(0)} \right)
\left(1-\tau_{\gamma_1}^{(0)} \right)
\left(1-\tau_{\gamma_0}^{(0)} \right) \right]\ \Gamma \cr}
\EQNO\sv{\zimmer} $$
A simple power counting argument shows that the last term in the above
equation is of order $1/m_w^4$ and will be discarded. The first, second
and third terms in Eq. (\zimmer) can be identified to the first,
second and third diagrams on the right in Fig. 6, respectively.
Note that Eq. (\zimmer) uses subtraction at zero momentum to renormalize
the four-quark operators. Also, each of the first three terms are finite:
the Dirac algebra can be done in four dimensions, and
the external fermions carry implicitly a factor ${\cal P}$.
Let us now rewrite Eq. (\zimmer) such that the four-quark
operators are renormalized in the MS scheme:
$$ \eqalign{
\Gamma =&\
\left(1-\tau_{\Gamma}^{(\epsilon)} \right)
\lim_{\epsilon \to 0}\;
\left(1-\tau_{\gamma_1}^{(\epsilon)} \right)
\tau_{\gamma_0}^{(0)} \ \Gamma \cr
&+
\left(1-\tau_{\Gamma}^{(\epsilon)} \right)
\lim_{\epsilon \to 0}\;
\left( \tau_{\gamma_1}^{(0)} - \tau_{\gamma_1}^{(\epsilon)} \right)
\left( 1 - \tau_{\gamma_0}^{(0)} \right)
\Gamma \cr
&+
\left( \tau_{\Gamma}^{(0)}-\tau_{\Gamma}^{(\epsilon)} \right) \cr
&
 \ \ \ \left[ 1 - \lim_{\epsilon \to 0}\;
\left( \tau_{\gamma_1}^{(0)} - \tau_{\gamma_1}^{(\epsilon)} \right)
\left( 1 - \tau_{\gamma_0}^{(0)} \right)
- \lim_{\epsilon \to 0}\;
\left( 1 - \tau_{\gamma_1}^{(\epsilon)} \right) \tau_{\gamma _0}^{(0)}\;
\right]
\Gamma \cr }
\EQNO\sv{\zimmerMS} $$
What we have done is simply to rearrange the terms in Eq. (\zimmer).
In the above equation, '' $\lim_{\epsilon \to 0} [\dots]$ '' means precisely
that we compute the $\underline{\rm finite\ subgraph}$
'' $[\dots]$ '' by first assuming its external particles are in four
dimensions. The result is then reinserted in $n$ dimensions.
The operation $\tau_{\gamma}^{(\epsilon)}$ extracts the pole
of the graph (or subgraph) $\gamma$ in the same fashion.
Let us recall that external particles carry a factor ${\cal P}$:
when the result is reinserted in $n$ dimensions, the factor ${\cal P}$
must be removed.

When this projector method is used to renormalize the various four-quark
operators, together with the use of an anti-commuting $\gamma ^5$, the
LLA results of $b \rightarrow s$ processes are identical to those of
Ref.[1].

\bigskip \noindent
{\bf 5. $b\rightarrow s$ processes in the LLA}

\medskip
In this section we discuss in the context of dimensional regularization
and evanescent operators the processes $b\rightarrow s$, in the $LLA$.
It is independent of section 4, and no projector is used.
Here we present in the $MS$ scheme
a two loop calculation of the mixing of all the evanescent operators
with the physical dimension five operators
(also called the dipole moment operators).
The results in other schemes can be obtained with the use of Eq. (\xth35).
We want to take note that we are now dealing with an infinite set of
operators.  It follows that the dimension of the mixing matrix is infinite.
In order to make mathematical sense, in principle one needs to
address the issues of boundedness, convergence, asymptotic completeness, etc.
These unfortunately we do not know how to deal with because some
of the matrix elements are in fact unbounded in value.  An example
of inconsistency is given to show the potential pitfalls.  On a
formal basis, however, we shall show explicitly that there is
a similarity transformation which will decouple the evanescent
set from the physical operators we are interested in.  Thus, based
on this 'physical argument', one may give meaning to the mixing
matrix as such.
\medskip
For $b\rightarrow s$ processes, the commonly used basis of physical
operators is
$$\eqalign{ \baselineskip 18pt
&  O_{61} =\, ({\bar{s}_{\alpha}}
{\gamma^{\mu}} {P_{\!\!{_L}}}
{c_\alpha} )
\,({\bar{c}_{\beta}}
{\gamma^{\mu}} {P_{\!\!{_L}}}
{b_\beta} ), \cr
& O_{62} =\, ({\bar{s}_{\alpha}}
{\gamma^{\mu}} {P_{\!\!{_L}}}
{c_\beta} )
\,({\bar{c}_{\beta}}
{\gamma^{\mu}} {P_{\!\!{_L}}}
{b_\alpha} ), \cr
& O_{63} =\, ({\bar{s}_{\alpha}}
{\gamma^{\mu}} {P_{\!\!{_L}}}
{u_\alpha} )
\,({\bar{u}_{\beta}}
{\gamma^{\mu}} {P_{\!\!{_L}}}
{b_\beta} ), \cr
& O_{64} =\, ({\bar{s}_{\alpha}}
{\gamma^{\mu}} {P_{\!\!{_L}}}
{u_\beta} )
\,({\bar{u}_{\beta}}
{\gamma^{\mu}} {P_{\!\!{_L}}}
{b_\alpha} ), \cr
& O_{65} =\, ({\bar{s}_{\alpha}}
{\gamma^{\mu}} {P_{\!\!{_L}}}
{b_\alpha} )
\,({\bar{u}_{\beta}}
{\gamma^{\mu}} {P_{\!\!{_L}}}
{u_\beta}
+\ldots
+{\bar{b}_{\beta}}
{\gamma^{\mu}} {P_{\!\!{_L}}}
{b_\beta} ), \cr
& O_{66} =\, ({\bar{s}_{\alpha}}
{\gamma^{\mu}} {P_{\!\!{_L}}}
{b_\beta} )
\,({\bar{u}_{\beta}}
{\gamma^{\mu}} {P_{\!\!{_L}}}
{u_\alpha}
+\ldots
+{\bar{b}_{\beta}}
{\gamma^{\mu}} {P_{\!\!{_L}}}
{b_\alpha} ), \cr
& O_{67} =\, ({\bar{s}_{\alpha}}
{\gamma^{\mu}} {P_{\!\!{_L}}}
{b_\alpha} )
\,({\bar{u}_{\beta}}
{\gamma^{\mu}} {P_{\!\!{_R}}}
{u_\beta}
+\ldots
+{\bar{b}_{\beta}}
{\gamma^{\mu}} {P_{\!\!{_R}}}
{b_\beta} ), \cr
& O_{68} =\, ({\bar{s}_{\alpha}}
{\gamma^{\mu}} {P_{\!\!{_L}}}
{b_\beta} )
\,({\bar{u}_{\beta}}
{\gamma^{\mu}} {P_{\!\!{_R}}}
{u_\alpha}
 +\ldots
+{\bar{b}_{\beta}}
{\gamma^{\mu}} {P_{\!\!{_R}}}
{b_\alpha} ), \cr
& O_{51} = {1 \over 2}i{g_s}\,{\bar{s}}
{G_{\mu \nu}} {\sigma_{\mu \nu}}
{({m_{s}\,P_{\!\!{_L}} + m_b\,P_{\!\!{_R}}})}b,  \cr
& O_{52} = {1 \over 2}ie{Q_{d}}\;{\bar{s}}
{F_{\mu \nu}} {\sigma_{\mu \nu}}
{({m_{s}\,P_{\!\!{_L}} + m_b\,P_{\!\!{_R}}})}b  \ \ \ \ ; \ \ \ \
Q_d=-{1\over3}. \cr} \EQNO\sv{\xbasis}  $$
$\alpha$ and $\beta$ are color indices.

\medskip
A few words about this choice of basis are in order. When the operators
$O_{65},\ \dots,\ O_{68}$ are inserted in the Green functions
corresponding to the processes $b\rightarrow s\gamma, s G$,
one encounters traces of Dirac matrices and $\gamma^5$. This has led
some authors to use a scheme different from the naive dimensional
regularization scheme (i.e. $\gamma^5$ anti-commuting), such as
the t'Hooft-Veltman scheme or the dimensional reduction scheme.
It should be observed that when we include evanescent operators,
this $\gamma^5$ problem is artificial and is created
by the choice of the basis listed in Eq. ({\xbasis}). One can in fact choose
a different basis of physical operators and use an anticommuting
$\gamma^5$ unambiguously.
For instance, one can replace the operators $O_{65},\ \dots,\ O_{68}$ by
the following new operators
$$\eqalign{\baselineskip=18pt
& Q_{65} =\, ({\bar{s}_{\alpha}} {\gamma^{\mu}} {P_{\!\!{_L}}} {b_\alpha} )
\;({\bar{u}_{\beta}} {\gamma^{\mu}} {u_\beta} + \ldots + {\bar{b}_{\beta}}
{\gamma^{\mu}} {b_\beta} ), \cr
& Q_{66} =\, ({\bar{s}_{\alpha}} {\gamma^{\mu}} {P_{\!\!{_L}}} {b_\beta} )
\;({\bar{u}_{\beta}} {\gamma^{\mu}} {u_\alpha} + \ldots + {\bar{b}_{\beta}}
{\gamma^{\mu}} {b_\alpha} ), \cr
& Q_{67} =\, ({\bar{s}_{\alpha}} {\gamma^{[\mu}} {\gamma^{\nu}}
{\gamma^{\sigma ]}} {P_{\!\!{_L}}} {b_\alpha} ) \; ({\bar{u}_{\beta}}
{\gamma^{[\mu}} {\gamma^{\nu}} {\gamma^{\sigma ]}} {u_\beta}
+ \ldots
+ {\bar{b}_{\beta}} {\gamma^{[\mu}} {\gamma^{\nu}} {\gamma^{\sigma ]}}
{b_\beta} ), \cr
& Q_{68} =\, ({\bar{s}_{\alpha}}
{\gamma^{[\mu}} {\gamma^{\nu}} {\gamma^{\sigma ]}} {P_{\!\!{_L}}} {b_\beta} )
\; ({\bar{u}_{\beta}}
{\gamma^{[\mu}} {\gamma^{\nu}} {\gamma^{\sigma ]}}
 {u_\alpha} + \ldots + {\bar{b}_{\beta}}
{\gamma^{[\mu}} {\gamma^{\nu}} {\gamma^{\sigma ]}} {b_\alpha} ),  \cr}
\EQNO\sv{\xbasisp}.  $$
which do not generate any trace involving $\gamma^5$.
Thus if we include the corresponding evanescent four-quark
operators (see below), then the use of a $\gamma^5$ anticommuting with all
the $\gamma^{\mu}$ is justified.

Following the notation of Ref. [8], we define the totally antisymmetric
matrices
$$ \eqalign{
& \G{\mu}{p} \DEF \gamma^{[\mu_1}\dots \gamma^{\mu_p]}, \cr}
\EQNO \sv{\fivethree}$$
%
%
with the understanding that $\G{\mu}{0}$ is
the identity matrix.
The basis of evanescent operators that can mix with $Q_{67,\,68}$
can be chosen as (below $p$ is odd, and $p\ge 5$):
$$\eqalign{
& Q_{67}^p =
\, ({\bar{s}_{\alpha}} \G{\mu}{p} {P_{\!\!{_L}}} {b_\alpha} )
\,({\bar{u}_{\beta}} \G{\mu}{p}  {u_\beta}
+ \ldots
+ {\bar{b}_{\beta}} \G{\mu}{p} {b_\beta} ), \cr
& Q_{68}^p =
\, ({\bar{s}_{\alpha}} \G{\mu}{p} {P_{\!\!{_L}}} {b_\beta} )
\,({\bar{u}_{\beta}} \G{\mu}{p}  {u_\alpha}
+ \ldots
+ {\bar{b}_{\beta}} \G{\mu}{p} {b_\alpha} ).  \cr}
\EQNO \sv{\xQev}  $$
We have checked explicitly that these are the only evanescent
operators that can mix with the dimension five
operators $O_{51,\,52}$, in the MS scheme.

Let us first rewrite Eq. (\xO3) as:
$$O_{\ii}^{ren} = \sum_{\jj} Z_{\ii\jj} \mu^{-\epsilon D_{\jj}}
 O_{\jj}^{bare}  \EQNO\sv{\fivefive}  $$
where the matrix $Z$ is dimensionless,
$D_{\ii}=0$ if ${\ii}$ corresponds to $O_{51,52}$, and $D_{\ii}=1$
if ${\ii}$ corresponds to a four quark operator (physical or evanescent).
The factor $\mu^{-\epsilon D_{\ii}}$ is introduced to make the mass dimension
of $O_{\ii}^{ren}$ constant: $dim\left(O_{\ii}\right)=n+2$ for all ${\ii}$.
With this choice, all the Wilson coefficients have dimension $mass^{-2}$.
To order $g_s^2$, we have in the MS scheme:
$$ \eqalign{
& Z = 1 + {g_s^2 \over 8\pi^2 \, \epsilon} B,  \cr
& \gamma_{\ii\jj} = - {g_s^2\over 8\pi^2}
\left(\, 1+D_{\ii}-D_{\jj} \,\right) B_{\ii\jj}. \cr}
\EQNO \sv{\fivesix} $$
We want to calculate the matrix $B$. As discussed, because we will not
encounter any traces involving $\gamma^5$, we will take for definiteness
$\gamma^5$ anti-commuting.
Also, the calculation will be done in the general linear covariant gauges
''\ ${-1\over 2\xi }(\partial _\mu G_\mu )^2$\ '' for the gluons.

\medskip
First, we insert the evanescent operators
in a four quark process: this gives the mixing of these operators
with themselves. There are 6 one loop diagrams which give:
$$\eqalign{
B_{Q_{67}^{p}\,,\,Q_{67}^{q}} =&
 {1\over 6} \,\delta_{p+2,q}
 + ( 8 - {32\over 3} \,p + {8\over 3} \,p^2 ) \,\delta_{p,q} \cr
& + ( - 5 \,p + {41\over 6} \,p^2 - 2 \,p^3 + {1\over 6} \,p^4 )
\,\delta_{p-2,q}, \cr
B_{Q_{67}^{p}\,,\,Q_{68}^{q}} =&
 - {1\over 2} \,\delta_{p+2,q}
 + ( 15 \,p - {41\over 2} \,p^2 + 6 \,p^3 - {1\over 2} \,p^4 ) \,
\delta_{p-2,q}, \cr
B_{Q_{68}^{p}\,,\,Q_{67}^{q}} =&
 - {1\over 4} \,\delta_{p+2,q}
 + ( 3 - 6 \,p + {3\over 2} \,p^2 ) \,\delta_{p,q} \cr
& + ( {15\over 2} \,p - {41\over 4} \,p^2 + 3 \,p^3
  - {1\over 4} \,p^4 ) \,\delta_{p-2,q}, \cr
B_{Q_{68}^{p}\,,\,Q_{68}^{q}} =&
 - {7\over 12} \,\delta_{p+2,q}
 + ( - 1 + {22\over 3} \,p - {11\over 6} \,p^2 ) \,\delta_{p,q} \cr
& + ( {35\over 2} \,p - {287\over 12} \,p^2 + 7 \,p^3
  - {7\over 12} \,p^4 ) \,\delta_{p-2,q}. \cr
} \EQNO \sv{\xBev} $$
The above results will be needed when we insert the evanescent operators
in $b\rightarrow s\gamma$ and $b \rightarrow sG$, to order $g_s^2$. They
have been obtained with the help of the following identities:
$$\eqalign{
& \gamma^{\alpha} \G{\mu}{p} \gamma^{\alpha} =
(-1)^p (n-2p) \G{\mu}{p}, \cr
& \gamma^{\alpha} \G{\mu}{p} \otimes \gamma^{\alpha} \G{\mu}{p} =
 \G{\mu}{p} \gamma^{\alpha} \otimes \G{\mu}{p} \gamma^{\alpha} = \cr
& \hskip2truecm \G{\mu}{p+1} \otimes \G{\mu}{p+1}
+ p(n+1-p) \G{\mu}{p-1} \otimes \G{\mu}{p-1}, \cr
& \gamma^{\alpha} \G{\mu}{p} \otimes \G{\mu}{p} \gamma^{\alpha} = \cr
&\hskip1.5truecm   (-1)^p \left(\G{\mu}{p+1} \otimes \G{\mu}{p+1}
 - p(n+1-p) \G{\mu}{p-1} \otimes \G{\mu}{p-1} \right), \cr
} \EQNO \sv{\xId} $$
which were also derived in Ref. [8].

Next, we insert $Q_{67,68}^p$ in $b\rightarrow s \gamma$ and $b\rightarrow sG$
to the lowest order. This is illustrated in Fig. 7. The results are finite,
and will be used (at the end of the calculation)
to ''renormalize away'' these evanescent operators; in other words,
we will redefine them such that they vanish at the lowest order
(for the $LLA$). For this, we define a finite matrix $A$
such that if $O_i$ is evanescent, then
$$ \Gamma(O_i+A_{ij}\,O_j)^{b\rightarrow s \gamma, \;sG} \rightarrow 0
\ \ \ \ {\rm when }\ \ \ \epsilon \rightarrow 0.
\EQNO \sv{\xdA} $$
The above equation defines $A$ at lowest order in the strong coupling
constant, and $O_j=O_{51,\,52}$.

\medskip
Before we present the results, we would like to give the
various identities which are used to simplify the $n$-dimensional
Dirac $\gamma$-algebra. We can easily derive, for arbitrary $p$,
$$ \G{\mu}{p} \G{\mu}{p} = (-1)^{ p(p-1)/2 } {n! \over (n-p)!}.
\EQNO \sv{\xsq}$$
For $p \ge 5$ and  $n\rightarrow 4$, Eq. (\xsq) becomes
$$ \G{\mu}{p} \G{\mu}{p}  = - 24\, \epsilon \;
(-1)^{ p(p+1)/2 } \; (p-5)! \ + \ {\cal O} \left(\epsilon^2\right).
\EQNO \sv{\xsqp} $$
More generally, we have (see also Ref.[11]):
$$ \G{\mu}{p} \G{\nu}{q} \G{\mu}{p} = { \G{\mu}{p}\G{\mu}{p}
\over \G{\alpha}{q} \G{\alpha}{q} } \  \beta_q^{(p)} \
\G{\nu}{q}, \EQNO \sv{\xbet}$$
where the parameters $\beta_q^{(p)}$ are given by the
following recursion relation:
$$ \eqalign{
& \beta_0^{(p)} = 1 \ \ \ \ , \ \ \ \ \beta_1^{(p)} = (-1)^p (n-2p), \cr
& \beta_q^{(p)} = (-1)^{p+q-1} (n-2p) \;\beta_{q-1}^{(p)}
\;+\; (q-1)(n-q+2) \;\beta_{q-2}^{(p)} \cr} \EQNO \sv{\xbetp} $$
For the present calculation, the values of $\beta_q^{(p)},$ for
$q=0,1,\ldots,5$ only, are needed. The above recursion
relation is obtained with the help of the following identity:
$$\eqalign{
& \G{\mu}{p} \otimes \G{\mu}{p} = \sum_{r=0}^p \; \alpha_r^{(p)} \;
\gamma^{\mu_1} \dots \gamma^{\mu_r} \;\otimes\;
\gamma^{\mu_r} \dots \gamma^{\mu_1};
\cr
& \alpha_0^{(0)} = \alpha_1^{(1)} = 1 \hskip1truecm , \hskip1truecm
\alpha_0^{(1)} = 0, \cr
& \alpha_r^{(p)} = 0 \ \ \ \ {\rm if}\ \ \ r<0 \ \ \ \
{\rm or\ if}\ \ \ r>p ,\cr
& \alpha_r^{(p+1)} = p(n+1-p) \alpha_r^{(p-1)} + (-1)^p \alpha_{r-1}^{(p)}. \cr
} \EQNO \sv{\xalpha} $$
Also, the following relation holds:
$$\eqalign{
& \beta_q^{(p)} = \sum_{r=0}^q \;\alpha_r^{(q)} \;
\left[ \,(-1)^p (n-2p) \,\right]^r\cr}. \EQNO $$

\bigskip
Our results for the matrix $A$ are, when $p\ge 5$:
$$\eqalign{
& A_{Q_{67}^p,O_{51}} =
  A_{Q_{67}^p,O_{52}} =
\,{1\over 3}\,
  A_{Q_{68}^p,O_{52}} =
 { {\rm h}(p) \over \pi^2}, \cr
& A_{Q_{68}^p,O_{51}} = 0, \cr}
\EQNO \sv{\xAlow} $$
where
$$ {\rm h}(p) \DEF (-1)^{p(p+1)/2} \; (p-5)! \; (p-1)(p-3). \EQNO \sv{\xhofp}$$

Next, we insert these evanescent operators in $b\rightarrow s\gamma$
and $b\rightarrow sG$ to order $g_s^2$. This is illustrated in Figs. 8,
9 and 10.
To extract the infinities of the two loop diagrams,
we need to calculate the double pole ($1/\epsilon^2$) only; this is because
the Dirac $\gamma$-algebra always generates an extra
$\epsilon$, as can be seen from Eq. (\xsqp). In the MS scheme, it is a well
known fact that the double pole of a two-loop diagram is a polynomial in the
external momenta and masses. One can therefore expand each diagram in
a power series in the external momenta and masses, to order $mass^2$.
The calculation is straightforward and is done with
the help of the symbolic manipulation program Schoonschip Ref.[12]. The results
for $\gamma_{ij}$ (see Eq. ({\xhofp})) are:
$$\eqalign{
& \gamma_{Q_{67}^p,O_{51}} =
{g_s^2 \over 8 \pi^4} \; {\rm h}(p) \;( -{40\over 3} + {28 \over 3}\,p
- {7\over 3}\,p^2 ), \cr
& \gamma_{Q_{68}^p,O_{51}} =
{g_s^2 \over 8 \pi^4} \; {\rm h}(p) \;( -2 +8\,p -2\,p^2 ) \cr
& \gamma_{Q_{67}^p,O_{52}} =
{g_s^2 \over 8 \pi^4} \; {\rm h}(p) \; ( -{40\over 3} +{ 64\over 3}\,p
-{16\over 3}\,p^2 )\cr
& \gamma_{Q_{68}^p,O_{52}} =
{g_s^2 \over 8 \pi^4} \; (-8)\, {\rm h}(p) \; .\cr
} \EQNO \sv{\xanom} $$

\medskip
Let us imagine that we solve Eq. (13), which formally has the solution
$$ C(\mu) = C(m_w) \;T \;{\rm exp}\left(\,
\int_{m_w}^{\mu} {d\mu \over \mu} \gamma(\mu) \,\right), \EQNO \sv{\xsol} $$
where $T$ stands for normal ordering of the matrices $\gamma(\mu)$.
For the LLA results, the mixing matrices
$\gamma(\mu)$ commute for
different $\mu$ and therefore the $T$-product can be omitted. Thus We obtain:
$$ C(\mu) = C(m_w) \;{\rm exp}\left(\,
{\mathaccent "7E{\gamma} \over 2\,\beta_{\!F}} \,\log(\eta_w) \,\right),
\EQNO \sv{\xsolp} $$
where
$$\eqalign{
& \gamma \DEF {g^2_s\over 8 \pi^2} \mathaccent "7E{\gamma}, \cr
& \eta_w \ \DEF \ {g^2_s(\mu)  \over g^2_s(m_w) }  \ \DEF\
1 + {\beta_F \over 16\pi^2}\, g_s^2(\mu)\, \log{m_w^2 \over \mu^2}. \cr
} \EQNO \sv{\xeta} $$
In the above equation, the only non-vanishing boundary conditions needed
in the LLA are $C_{O_{61,63}}(m_w)$ and $C_{O_{51,52}}(m_w)$
(see Eq. (\xbasis)).

\medskip
The matrix $\mathaccent "7E{\gamma}$ has entries that become
large as $p$ becomes large (see Eqs. (\xBev \ and \ \xanom). This tells us that
the evanescent Wilson coefficients may become large as $p\rightarrow \infty$,
and the convergence of
$$\eqalign{
& \sum_{i=evan} C_{O_i}\,\gamma_{O_i,O_{51}},  \cr} $$
is immediately brought into question.  For example, we do not know how
to evaluate numerically $C_{Q_{67}^p}$ at say $\mu=5 \ {\rm GeV}$.

\medskip
The handling of such an infinite set of operators is quite delicate.
To illustrate, suppose one chooses instead a new basis $\{U_i\}$ of
evanescent operators, defined by:
$$ \eqalign{
& U_{67}^5 \DEF Q_{67}^5 \ ,\ \ \ \ \ U_{68}^5 \DEF Q_{68}^5, \cr
& U_{67}^{p+2} \DEF Q_{67}^p + {1 \over (p+1)(p-4) }\, Q_{67}^{p+2}
\ \ \ {\rm for\ }\ p \ge 5, \cr
& U_{68}^{p+2} \DEF Q_{68}^p + {1 \over (p+1)(p-4) }\, Q_{68}^{p+2}
\ \ \ {\rm for\ }\ p \ge 5. \cr}
\EQNO \sv{\ubasis} $$
This new basis has the property that in the limit
$\epsilon \rightarrow 0$, the renormalized Green
functions of $U_{67,68}^p$, $ p\ge 7$, vanish at the lowest order for
$b\rightarrow s \gamma$ and $b\rightarrow sG $. Only $U_{67,68}^5$ have
a non-vanishing value. Because the basis $\{U_i\}$ is infinite,
one may querry whether the operators $U_{67,\,68}^5$ are
in fact linearly independent of the other $U_{67,\,68}^p,\ \;p\ge 7$.
One may be flippant and formally writes
$$ U_{67}^{(5)} = \sum_{p=5}^{\infty} \, d_p\, U_{67}^{p+2} \EQNO \sv{\xqu}$$
where the coefficients $d_p$ are obtained by the following recursion relation:
$$ d_{p+2} = - {1 \over (p+1)(p-4)}\, d_p \ \ \ \ ;\ \ \ \ d_5=1.
\EQNO \sv{\xdp}$$
Is one allowed then to choose a {\it reduced basis}
$\{U_i\}$ which does not contain
the two evanescent operators $U_{67,\,68}^5$?
Formally, it seems that nothing should prevent one to start with this
reduced new basis to renormalize all the physical and evanescent operators.
The question is then: do the full basis and the reduced basis give the
same on-shell results? If one uses Eq. (\xqu), then
the Green functions to $b\rightarrow s \gamma $ of $U_{67}^5$ at the
lowest order are not well defined,
since the term on the left in Eq. (\xqu) gives non-zero, and each term
on the right gives formally zero.
This ''problem'' is obviously related to the
fact that one has  an infinite number of evanescent operators.  It would
not exist if the number of evanescent operators were finite.
Unfortunately, as we have seen,
one cannot carry out the program of renormalization
with a finite number of evanescent operators, except for
a prescription in which no evanescent operator is introduced altogether.

\medskip
One may skirt this potential difficulty by the following:
the LLA result for the amplitude
of $b\rightarrow s \gamma$ is
$$ \left[ C_i \Gamma(O_i) \right]^{^{LLA}} = C_i^{^{LLA}} \;
\Gamma(O_i)^{lowest \ order}  \EQNO \sv{\xreslla} $$
where the sum is over {\it all} physical and evanescent operators.
If we use Eq. (\xdA), then we can rewrite Eq. (\xreslla) as
$$ \left[ C_i \Gamma(O_i) \right]^{^{LLA}} = \sum_{i=phys}
\left[ C_i^{^{LLA}} \; - \sum_{j=evan} C_j^{^{LLA}} \, A_{ji} \right]
\Gamma(O_i)^{lowest \ order}  \EQNO \sv{\xresllap} $$
The above result was obtained in the MS scheme. It is however
the same in all schemes (including mass-dependent schemes).
One has
redefined the evanescent operators such that their physical matrix
elements vanish at the lowest order.
This leads to
$$ \eqalign{
Z \rightarrow Z' &= 1 + A + {g_s^2 \over 8\,\pi^2\,\epsilon} B', \cr
		 &= (1+A)\,Z. \cr} \EQNO \sv{\xzprime} $$
This manipulation was implicitly used in Ref. [2,3] and the anomalous
dimension matrix becomes (see Eq. (\xth35)):
$$ \gamma \rightarrow \gamma' = (1+A)\,\gamma\,(1-A), \EQNO \sv{\xnewM} $$
where $\gamma$ is the mixing matrix obtained in the MS scheme.
To the order, we have used $\ (1+A)^{-1}=1-A\ $, $A^2=0$.  One can easily
check that
$$ \eqalign{
& \gamma '_{Q_{67}^p,O_{51}} \ = \ 0, \cr
& \gamma '_{Q_{68}^p,O_{51}} \ = \ 0, \cr
& \gamma '_{Q_{67}^p,O_{52}} \ = \ 0, \cr
& \gamma '_{Q_{68}^p,O_{52}} \ = \ 0. \cr}
\EQNO \sv{\expl} $$
Effectively, the mixing matrix between the four quark and the dipole
moment operators is collapsed into a finite dimensional one,
which has elements from physical to physical.
\bigskip
We can now finish constructing the mixing matrix between the physical four
quark operators and the dipole moment operators.  They agree with the
results by Ciuchini et. al.  Parenthetically, we remark again that in
view of our previous discussion this requires the assumption that
$\lim_{\epsilon \to 0}\sum_{i=evanes}C_i\Gamma(O_i)=0.$ In our
opinion, this has not been firmly established because, as we have
seen, $C_{Q_{67}^p}$ may become unbounded as p approaches infinity.

\bigskip
The renormalized physical operators in
the above scheme (where evanescent operators  are first introduced then
renormalized away) are related to those in the naive scheme
in which there are no such operators by a finite transformation
which depends explicitly on $\mu$:
$$[O_{phys}]^{ev. \ scheme}=
\left( 1+g_s^2\log (\mu/ m_{light}) \ F + \dots \right)
[O_{phys}]^{naive \ scheme},
\EQNO\sv{\naive}$$
where the finite matrix $F_{ij}\sim\gamma_{ik}A_{kj}$.
This means that the Wilson coefficients cannot be of the form
given by Eq.(\formC) in both schemes.  One of the schemes must be
mass-dependent.  This is what we referred to at the beginning of
this article as the light-mass dependence of the $C_i$.  Thus, the
problem has been shifted to a problem of factorization, namely,
finding a prescription in which the
Wilson coefficients do not depend on the masses of the light particles.
This is our present understanding to account for the difference in
results between Ref.[1] and Ref.[3].
One has to go to at least three loop order to settle the issue.

\bigskip\noindent
{\bf 6. Concluding Remarks}

\medskip
In this article, we have attempted to locate the cause for the
difference in the short distance analysis of $b\rightarrow s
\gamma $ between the scheme in Ref.[1,4] in which no evanescent
operators
are introduced and those schemes in Ref.[2,3], which contain such
operators.
We have argued that if one follows a systematic formulation to arrive
at an effective Lagrangian, which for our preference is the
Zimmermann
rearrangement, then there is ample freedom in divising different
schemes.  We have explained how the Wilson coefficients $C_i$ depend
on the scheme used to renormalize the operators.  The boundary
conditions
also depend on the procedure.  This can be easily understood if we
observe that the identification and the subtraction of the "infinities"
are done differently in different schemes. Consequently, the $C_i$ to
that order must be different, which affect the matching conditions.
Such difference will exactly cancel out the difference in the
anomalous
mixing between various schemes.  It is the S-matrix elements of the
product $\sum C_i O_i$ which is independent of the scheme used.
We have also argued that the problem of $\gamma^5$ is artificial,
since all quantities are finite to begin with; provided one is consistent
all prescriptions for handling $\gamma^5$ are equivalent.

In section 4, we have elaborated on a version of the t'Hooft
Veltman scheme, in which all external particles are kept in four
dimensions and we have developed an algorithm to handle the Dirac
algebra for processes with more than two fermions, in particular
when four-quark operators are introduced into the analysis.
This scheme as laid out so far may, however, be incomplete,
since it breaks Lorentz invariance in $n$ dimensions.
It would be instructive to construct a Lagrangian which is Lorentz
invariant only in four dimensions; for instance, we could replace
every fermion field $\psi$ by $\cal{P}\psi$
(and $\bar{\psi}$ by $\bar{\psi}\cal{P}$) \underbar{at the lagrangian level}.
This would have the advantage that
in the evaluation of Feynman graphs derived from such a Lagrangian,
the Dirac $\gamma$-algebra could be handled in four dimensions. We do not
know if such a theory (or scheme) is consistent, but from a theoretical
point of view we believe it is worth investigating.

In section 5, we have followed the approach in Ref.[2,3] in
introducing evanescent operators.
We have shown that some entries in the mixing
matrix are unbounded.
However, by a {\it formal finite} transformation, which is induced by imposing
a requirement that the matrix elements of the evanescent operators
vanish at the lowest order, the mixing matrix
elements of the new evanescent basis with the magnetic operators
formally vanish also. The results of Ref.[3] (as far as the anomalous
dimension mixing matrix of the physical operators) have been reproduced.
We nonetheless are concerned that this scheme lacks certain justification
in some formal manipulations which are related to
the convergence of the infinite series with coefficients that
eventually go to infinity.

We have also stressed that in going from finite order analysis to
LLA with the help of RGE, we need to make sure that mass independence
of $C_i$ is valid to facilitate the summation of leading logarithms.
By this, we mean that the only $\log \mu$ dependence in $C_i$ can
only be in the form $\log (M_{heavy}/\mu )$. We have not been able to
establish this property in either scheme.
We however believe that this is the most important aspect
which needs to be clarified: it is our contention that
this should account for the difference between the results of Ref.[1]
and Ref.[3] and an unambiguous way to settle this is to perform an
explicit higher loop Feynman diagram calculation, where the
discrepancies first show up.  Further work needs to be done.

\bigskip\noindent
{\bf Acknowledgement}
\medskip
We are grateful to M.~Veltman for sharing his insight.  This work has been
partially supported by the U. S. Department of Energy.

\vfill\eject

\noindent
{\bf References}
\bigskip

\settabs 20 \columns
\+ [1] & K. Adel and York-Peng Yao,  Modern
             Physics Letters A {\bf 8}, 1679 (1993). \cr

\medskip
\+ [2] & M. Misiak,   Nucl. Phys. {\bf B393}, 23 (1993); some of the earlier
                 works are:\cr
\+     & B.~Grinstein, R.~ Springer and M.~B.~Wise,  Phys. Lett.
                 {\bf 202 B}, 138 (1988); \cr
\+     & P.~Cho and B.~Grinstein,  Nucl. Phys.
                 {\bf B365}, 138 (1991); \cr
\+     & R.~Griganis, P.~J.~O'Donnell, M.~Sutherland and H.~Navalet,
                  Phys. Lett. {\bf 213 B}, \cr
\+     & 355 (1988); R.~Griganis, P.~J.~O'Donnell and M.~Sutherland,
          Phys. Lett. {\bf 237 B};  \cr
\+     & G.~Cella, G.~Curci, G.~Ricciardi and A. Vicer\'e,  Phys. Lett.
            {\bf 248 B}, 181 (1990). \cr
\medskip
\+ [3] & M. Ciuchini, E.Franco, G. Martinelli, L. Reina, L. Silvestrini,
               LPTENS 93/28, \cr
\+     & ROME 93/958, ULB-TH 93/09. \cr

\medskip
\+ [4] & K. Adel and York-Peng Yao, Phys. Rev. D{\bf 49}, 4945 (1994).\cr

\medskip
\+ [5] & Y.~Kazama and  Y.-P.~Yao,  Phys. Rev. {\bf D 21}, 1116 (1980).\cr

\medskip
\+ [6] & W.~Zimmermann, in {\sl Lectures in Elementary Particles and Quantum
        Field} \cr
\+     &  {\sl Theory}, edited by S.~Deser {\sl et al.}
       (MIT Press, Cambridge, Mass., (1971),\cr
\+    & Vol. I, p.397.\cr

\medskip
\+ [7] & A. Buras and P. Weisz, Nucl. Phys. {\bf B333}, 66 (1990). \cr

\medskip
\+ [8] & M.J. Dugan and B. Grinstein,
             Phys. Lett. {\bf 256 B}, 239 (1991). \cr

\medskip
\+ [9] & M.~Veltman, UM-TH-88-17. \cr

\medskip
\+ [10] & J.~Collins, in {\sl Renormalization},
 (Cambridge U.~Press, Cambridge, 1971).\cr

\medskip
\+ [11] & It has been brought to our attention that identities
     similar to Eq.({\xbet })\cr
\+      &  were derived independently in
       Giancarlo~Cella, Giuseppe~Curci,\cr
\+      & Giulia~Ricciardi and Andrea~Vicer\' e,
        IFUP-TH 9/94, HUTP-94/A001.\cr
\medskip
\+ [12] & SCHOONSCHIP program by M.~Veltman, unpublished. \cr

\vfill \eject

\noindent
FIGURE CAPTION

\medskip
\noindent
Figure 1. Diagrams contributing to $b\rightarrow s \gamma$ at lowest order
in the Standard Model.

\medskip
\noindent
Figure 2. The Zimmermann rearrangement for a one-loop graph contributing to
$b\rightarrow s \gamma$.

\medskip
\noindent
Figure 3. The Zimmermann rearrangement for
four-quark processes at the tree level. The effective
vertex on the right is reinserted in one-loop diagrams contributing
to $b\rightarrow s \gamma$, as shown on the first graph on the right in Fig. 2.

\medskip
\noindent
Figure 4.  An example of the Zimmermann rearrangement for
four-quark processes to order $g_w^2 g_s^2$.

\medskip
\noindent
Figure 5. Graphical representation of the next-to-leading
logarithmic approximation final results for four-quark processes.
In this approximation we are summing all terms of the forms
$g_s^{2n} \log^n (m_w^2/\mu^2)$ and
$g_s^{2n+2} \log^n (m_w^2/\mu^2)$. The final results are prescription
independent since the "ambiguities" (concerning $\gamma^5$ and the inclusion
or exclusion of evanescent operators) in the graphs on the right
exactly cancel.

\medskip
\noindent
Figure 6. Example of the Zimmermann rearrangement to order $g_w^2 g_s^4$ for
$b\bar{c} \rightarrow s \bar{c}$. In the LLA, one keeps the third graph on
the right only; in the NLLA, one keeps the second graph on the right as well;
in the next-to-next-to-leading logarithmic approximation, one must also
include the contribution of the first graph on the right.
Again, all the ambiguities we encounter in
the graphs on the right cancel exactly, since the graph on the left is
finite and prescription independent.

\medskip
\noindent
Figure 7. Contribution of four-quark operators to $b\rightarrow s \gamma, \ sG$
at lowest order. These diagrams also give the finite mixing
matrix $A$ which is used to "renormalize away" the evanescent operators.

\medskip
\noindent
Figure 8. Two-loop graphs contributing to the mixing of four-quark operators
with the dipole moment operators. The external wavy line represents a photon
or a gluon, and the black square represents an insertion of a
four-quark operator.

\medskip
\noindent
Figure 9. More two-loop graphs contributing to $b\rightarrow sG$ with an
insertion of a four-quark operator.

\medskip
\noindent
Figure 10. Graphs needed to remove the subdivergences of the two-loop
graphs in Figs. 8,9. The crosses represent the necessary one-loop
counterterms.

\vfill\eject


\end